\documentstyle[preprint,aps]{revtex}
\begin{document}
\draft
\title{Onset of Collectivity in Neutron Deficient $^{196,198}$Po}

\author{L. A. Bernstein,$^1$ J. A. Cizewski,$^{1}$ H.-Q.
Jin,$^1$\cite{A} W. Younes,$^{1}$ R. G. Henry,$^{1,2}$\cite{B}
L. P. Farris,$^{1,3}$ A. Charos,$^{1}$ M. P.
Carpenter,$^2$ R. V. F. Janssens,$^2$
T. L. Khoo,$^2$ T. Lauritsen,$^2$ I. G.
Bearden,$^{2,5}$\cite{C} D. Ye,$^{2,6}$\cite{D}
J. A. Becker,$^3$ E. A.
Henry,$^3$ M. J. Brinkman,$^3$\cite{A}
J. R.  Hughes,$^3$ A.
Kuhnert,$^3$\cite{E}
T. F. Wang,$^3$ M. A.
Stoyer,$^{3,4}$ R. M.
Diamond,$^4$  F. S. Stephens,$^4$ M. A.  Deleplanque,$^4$ A. O.
Macchiavelli,$^4$ I. Y.  Lee,$^4$ B. Cederwall,$^4$
J. R. B. Oliveira,$^4$\cite{F}
J. Burde,$^4$ P. Fallon,$^4$
C. Duyar,$^7$ J. E. Draper,$^7$ E. Rubel,$^7$ D. T.
Vo,$^8$}

\address{
$^1$  Rutgers University, New Brunswick, NJ  08903  USA\\
$^2$ Argonne National Laboratory, Argonne, IL  60439 USA \\
$^3$ Lawrence Livermore National Laboratory, Livermore, CA  94550 USA\\
$^4$ Lawrence Berkeley Laboratory, Berkeley CA  94720 USA\\
$^5$ Purdue University, West Lafayette, IN  47907 USA\\
$^6$  Notre Dame University, Notre Dame, IN  46556  USA\\
$^7$ University of California , Davis,  CA  95616  USA\\
$^8$ Iowa State University, Ames, IA  50011 USA}

\date{\today}
\maketitle
\begin{abstract}
We have studied via in-beam $\gamma$-ray spectroscopy $^{196}$Po and
$^{198}$Po, which are the first neutron-deficient Po isotopes to
exhibit a collective low-lying structure.  The ratios of
yrast state energies and the E2 branching ratios of
transitions from non-yrast to yrast states are indicative of
a low-lying vibrational structure.  The onset of collective
motion in these isotopes can be attributed to the opening of
the neutron i$_{13/2}$ orbital at N$\approx$112 and the resulting large
overlap between the two valence protons in the h$_{9/2}$ orbital
and the valence neutrons in the i$_{13/2}$ orbital.
\end{abstract}

\narrowtext

\section{Introduction}

The polonium isotopes, with two valence protons beyond the closed Z=82
core, provide an excellent laboratory in which to study the transition
between single-particle and collective behavior in a nuclear system.
The low-lying structure and energy spacings in $^{210}$Po  with N=126
can be
described by two h$_{9/2}$ protons in spherical shell models.  As the
number
of neutrons decreases, the large number of valence particles makes a
shell model description of the low-lying structure less meaningful and
the onset of a collective structure in the low-lying states is expected.

However, the type of collective motion that occurs is an open question.
One possibility, due to the small number of valence protons, is a
vibrational spectrum.  Vibrational systems are characterized by equal
energy spacings, nearly degenerate phonon multiplets, and
$\Delta$N$_{ph}$ = $\pm$ 1 E2
selection rules, where N$_{ph}$  is the number of phonons for that
state.   A second possibility has been suggested in earlier
studies~\cite{1,2} of the Po nuclei.  As the middle of the neutron
shell is approached, a 4 particle-2 hole (4p-2h) excitation, in
particular the ($\pi$h$_{9/2}$)$^4$($\pi$s$_{1/2}$)$^{-2}$
excitation, could play an important role in Po nuclei, just as
evidence for 2p-2h proton excitations is observed in the Pb
isotones~\cite{2}.  The 4p-2h configuration has a larger effective
proton valency and is expected
to be moderately deformed.  One signature of a 4p-2h structure is strong
E0 transitions to the ``normal'' ($\pi$h$_{9/2}$)$^2$ structure, which
occurs when
there is mixing between ``normal'' and 4p-2h configurations.  The exact
form of collectivity and the degree to which it persists as a function
of excitation energy and angular momentum will be strongly influenced by
the nature of the orbitals located near the Fermi surface, which are
displayed in Fig.~1.  As we shall show, the unique-parity
$\nu$i$_{13/2}$ orbital
plays a major role in the collectivity as the more neutron-deficient
nuclei near N $\approx$ 112 are considered.

\section{Experiments}

We have performed two measurements of $^{196}$Po.  Initially, the
$^{172}$Yb($^{28}$Si,4n) reaction was studied at the Argonne Tandem - Linear
Accelerator System (ATLAS) facility, with beam energies of 141 and 145
MeV, using an enriched $^{172}$Yb target ($\approx$1mg/cm$^2$) on a
$\approx$10 mg/cm$^2$ Pb
backing.  The Argonne-Notre Dame BGO facility was used.  This array
consists of 12 Compton-suppressed Ge detectors and a 50-element
Bismuth-germanate (BGO) inner array.  Approximately 35 million
$\gamma$-$\gamma$
events, with
an inner array multiplicity K$\geq$2 and at least 2 suppressed Ge detectors,
were recorded to tape.  The data were sorted off-line into several
matrices, of which the one with K$\geq$5 was used for the majority of the
analysis.  This minimized the contributions from Coulomb excitation and
particle transfer channels in the data.  An example of a coincidence
spectrum is presented in Fig.~2, which displays the spectrum gated on
the 463-keV $2^+ \rightarrow 0^+$ transition.  Spin assignments were
determined
through the use of Directional Correlations of gamma rays from Oriented
nuclei (DCO) ratios.  The definition of the DCO ratios for the Argonne
$^{196}$Po data is given in Ref.~\cite{4}.  Table~1 summarizes the results
for the $\gamma$
rays in $^{196}$Po determined from the ANL data set; complete details of the
analysis are given in Ref.~\cite{5}.  The 5$^-$ spin-parity assignment to
the 1802-keV level is based on systematics and the stretched dipole
character of the 414-keV line.

In a second experiment at the 88'' Cyclotron at Lawrence Berkeley
Laboratory, $^{196}$Po was also produced with the same reaction and a 142 MeV
$^{28}$Si beam.  The target consisted of two thin
($\approx$500$\mu$g/cm$^2$) $^{172}$Yb foils.
The High Energy Resolution Array (HERA), which consisted of 20
Compton-suppressed Ge detectors and a 40 element BGO inner ball, was
used for
spectroscopy.  Approximately 65 million doubly-coincident events were
recorded to tape.  The results of the initial analysis confirmed the
level scheme obtained from the ANL experiment.  A ``triples matrix'' was
then constructed from all triple and higher fold events; for each sorted
event two $\gamma$ rays assigned to $^{196}$Po from the ANL data set
were required.
Pairs of coincidence gates were used to place a higher spin, non-yrast
band which could not be separated from competing channels in the ANL
data.  The level spectrum of $^{196}$Po deduced from both of our data
sets is displayed in Fig.~3.

Also at the LBL 88'' cyclotron $^{198}$Po  was studied via the
$^{174}$Yb($^{29}$Si,5n)
reaction with 141 and 146 MeV beams and the HERA spectrometer.  Both
thin (three stacked $\approx$500$\mu$g/cm$^2$ foils) and thick
($\approx$1mg/cm$^2$ on $\approx$12 mg/cm$^2$
Au backing) enriched $^{174}$Yb targets were used.  The data were sorted into
two matrices, with cuts on the inner-ball fold of K$>$6 and K$>$11.  A
normalized subtraction of these matrices allowed for separation of the
5n ($^{198}$Po) and 4n ($^{199}$Po) channels.  An extensive level scheme up
to E$_x$ $\approx$ 5.1 MeV
and J$\approx$20$\hbar$ was deduced and will be reported on
elsewhere~\cite{5}.  Concurrent
results~\cite{6}  by M. Lach {\it et al.}, on levels in $^{198}$Po
have been recently reported.

\section{Results}

\subsection{$^{196}$Po}

Earlier work by Alber {\it et al.}~\cite{1}, studied the delayed
$\gamma$ rays
depopulating the 850-ns 11$^-$ isomer in $^{196}$Po.  As summarized in
Fig.~3,
this level scheme has been extended via the first in-beam spectroscopy
measurements in this nucleus and spins and probable parities of the
levels have been determined using the ATLAS data.  Most of the results
of the isomer decay study have been confirmed in the present study, with
the exception of the proposed direct decay from the isomer to the
negative-parity states.  The 552- and 198-keV lines proposed in the
decay study of Ref.~\cite{1} were not observed.
Since both the 237- and 253-keV
transitions, and not the 552-keV $11^- \rightarrow 8^+$ transition
which directly
depopulates the isomer, are clearly evident in Fig.~2, it can be
inferred that both the 237- and 253-keV lines must be prompt and neither
of them depopulates the isomer directly.

Our prompt spectroscopy has also yielded two new branches which bypass
the 11$^-$ isomer.  The first cascade consists of 4 transitions up to
J=13$\hbar$
and is connected to the yrast 6$^+$ and 8$^+$ states via two $\gamma$
rays, 584 and
652 keV, respectively.  The second extension is a sequence of 3 mutually
coincident transitions feeding into the 7$^-$, 9$^-$ and (10$^+$)
states.  This
cascade was found in the LBL data in spectra that were double-gated on
known low-lying transitions in $^{196}$Po.  Since this band could only be
identified using a sum of doubly-coincident gates on the 237, 253, 414,
and 529 keV lines, no multipolarity information for this cascade could
be determined.  The new lines did not have sufficient intensity to
extract DCO ratios from the ANL DCO data.  We have also added two new
levels above the 9$^-$ state at 2292 keV.

The $2_2^+$ and $4_2^+$ states are of particular interest in the
low-lying level
scheme.  They feed  into the yrast cascade via the 396- and 497-keV
$\Delta$J=0
transitions, respectively.  We summarize in Table 2 the relative
intensities for transitions in the $4_2 \rightarrow 4_1 \rightarrow
2_1$ and the $2_2 \rightarrow 2_1 \rightarrow 0$ cascades.
These values were determined from ANL spectra gated on the 414- and 529-
keV lines which directly feed the $4_2^+$ and $2_2^+$ levels, respectively.
Within the statistical uncertainties the intensities of the transitions
in these cascades are identical and there is no indication in either of
our data sets of any missing $\gamma$-ray intensity in the $\Delta$J=0
transitions, and hence no measureable evidence for E0 components.  This is in
contrast to the analysis of Ref.~\cite{1}, which claimed to see intensity
imbalances, although no supporting data were presented.

\subsection{$^{198}$Po}

The nucleus $^{198}$Po  was previously investigated in Refs.~[1,6];
the level scheme we have deduced for the low-lying
levels agrees with the earlier work.  The
low-lying level spectrum is very similar to that of $^{196}$Po in that
there is
an yrast cascade of similarly spaced $\gamma$ rays, as well as
non-yrast $2_2^+$ and
$4_2^+$ levels which feed into the yrast branch via $\Delta$J=0
transitions.  We
summarize in Table 2 the relative intensities for transitions in the
$4_2 \rightarrow 4_1 \rightarrow 2_1$ and the $2_2 \rightarrow 2_1
\rightarrow 0$ cascades in $^{198}$Po.  These values were
determined from spectra gated on the 391- and 444-keV lines which
directly feed the $4_2^+$ and $2_2^+$ levels, respectively.   As with
$^{196}$Po, no
clear evidence was found for missing intensity in the $\Delta$J=0
transitions, and hence no evidence for measureable E0 components.
This result is in
contrast to previous~\cite{1} measurements of $^{198}$Po. The analysis
reported in
Ref.~\cite{1}  indicated $\alpha_{\rm tot}$ ($2_2 \rightarrow 2_1$
transition) $\geq$2 and most likely $\alpha_{\rm tot} \approx$ 5.
Our data show no statistically significant intensity imbalance; given
our error bars we extract $\alpha_{\rm tot} <$2.  The $\alpha_{\rm
tot}$ = 0.52(23) for the $2_2 \rightarrow 2_1$
line deduced in Ref.~\cite{1} is small and not that different from the expected
value for an M1 transition.  Again our data show no statistically
significant intensity imblance and within our
error bars $\alpha_{\rm tot} <$0.6.

\section{Discussion}

The systematics of the even-even Po nuclei is displayed in Fig.~4.  The
heavier isotopes (A$>$200) have remarkably similar yrast 2$^+$, 4$^+$,
6$^+$ and 8$^+$
energy spacings, with low-lying 2$^+$ states and closely spaced 6$^+$
and 8$^+$
levels.  However, a change is evident in the vicinity of A$\approx$198.   A
simple way of examining the low-lying structure of a nucleus is to
consider the ratio of the yrast state energies.  $^{210}$Po, which is
semi-magic with N=126, has a ratio of the yrast 4$^+$ and 2$^+$ energies,
R(4/2) = 1.21.  This is precisely what is predicted for a two-particle system
with a residual surface-delta interaction~\cite{9}.   In a vibrational
system, the expected value for R(4/2) is 2.0.  All of the Po nuclei from
$^{206}$Po$_{122}$ to $^{196}$Po$_{112}$ exhibit R(4/2) values close
to 2.  This is a strong
indication that the states up to J=4 have dominant vibrational character
almost as soon as the neutron shell is opened.  On the other hand, the
close spacing between the 6$^+$ and 8$^+$ states suggest dominant
($\pi$h$_{9/2}$)$^2$ components.

The transition to collectivity for the higher spin states (J $>$ 4) occurs
deeper in the neutron shell, near N=112.  Two features are distinctive
in the low spin level schemes of $^{196}$Po and $^{198}$Po, as seen
in Fig.~4.
The first is the almost equal spacing between the yrast 0$^+$, 2$^+$,
4$^+$ and 6$^+$ levels.  The second is the non-yrast  $4_2^+$ and
$2_2^+$  states, which in
$^{196}$Po are nearly degenerate with the yrast 6$^+$ and 4$^+$
states, and form
structures which are reminiscent of 3- and 2-phonon multiplets,
respectively.  In addition, $^{198}$Po  is the first nucleus in the isotopic
chain to show a decrease in the yrast 6$^+$ energy.  In $^{196}$Po the
6$^+$ energy
has dropped an additional 327 keV to a point located almost exactly
half-way between the yrast 4$^+$ and 8$^+$ states.  These features combine to
form a low-lying level scheme in $^{196}$Po that is a classic example of a
vibrational structure.

The yrast energy ratios for $^{196,198,200}$Po are summarized in Table 3.  In
Po the valence protons are in the h$_{9/2}$ orbital, with a maximum coupled
angular momentum of 8.  Therefore, to examine the limits of low-lying
collective motion in Po, the ratios of the higher spin yrast levels,
such as R(6/4) and R(8/6) should be considered.  The predictions of
energy ratios for ($\pi$h$_{9/2}$)$^2$, harmonic vibrational, and
rotational models
are also listed in Table 3.  The data for both $^{196}$Po and $^{198}$Po  are
clearly best reproduced by the vibrational expectations, even up to the
R(8/6) ratio in $^{196}$Po.  In contrast, in $^{200}$Po the energy
ratios for the
higher spin states favor a two-proton configuration.  The difference in
the collectivity of the 6$^+$ states in $^{198}$Po  and $^{200}$Po is
also supported
by the absolute B(E2) values.  The B(E2; 8$^+ \rightarrow$ 6$^+$)
value~\cite{6}  in $^{198}$Po  is
significantly smaller than in the heavier isotopes, which indicates a
structural difference between these 8$^+$ and 6$^+$ states, and probably a
more collective nature for the 6$^+$ level in $^{198}$Po.

The branching ratios of transitions depopulating the low-lying states
also probe the nature of the excitation.  If the low-lying structure in
$^{198,196}$Po is vibrational, then the selection rule
$\Delta$N$_{ph}$  = $\pm$1 for E2
transitions between phonon multiplets should be valid.  Although no
absolute B(E2) values were obtained in the present measurements, the
ratio of B(E2) values can be calculated from the ratio of the $\gamma$-ray
intensities of the two depopulating transitions as seen in a coincidence
gate placed on a $\gamma$-ray feeding the level of interest.  The
assumption is
made that the observed $\Delta$J=0 transitions are of pure E2 character.  This
is a reasonable assumption because in the Pb isotones the analogous
transitions have small M1 components~\cite{2}. The comparisons between
experimental values and the theoretical predictions for a vibrational,
rotational, and a 4p-2h configuration are given in Table 4 for both
$^{198}$Po  and $^{196}$Po.  For the vibrational limit, the branching
ratios are
based on the coefficients of fractional parentage, which govern the wave
functions of the members of the phonon multiplets.  For the rotational
limit, transitions out of the band are forbidden.  Typically, these
``forbidden'' transitions are 20-50 times weaker than the allowed
branches.   If the non-yrast states are members of a 4p-2h
configuration, then, to first-order, they will not interact with the
normal two-proton configurations via the one-body E2 or M1 operators.
However, E0 transitions are expected if there is any mixing between the
spherical 2p, and more deformed 4p-2h, configurations, due to the
difference in their radii.  In the present data, E0 components in the
transitions between the ``normal'' and 4p-2h states would be signaled by
missing gamma-ray intensity in the $\Delta$J=0 transitions.  Large E0
components ($\rho^2{\rm (E0)} \leq {\rm 7 x 10}^2$) are seen in the
decay of the $2_2^+$ states in
the Pb isotones~\cite{2}.   In contrast, our data, summarized in Table 2, show
no clear evidence for missing intensity in any of the $\Delta$J=0 transitions
depopulating the non-yrast $4_2^+$  and $2_2^+$  states in either
$^{198}$Po  or $^{196}$Po.
As displayed in Tables 3 and 4, the relative B(E2) ratios of the
non-yrast transitions, together with the energy spacings, are only
consistent with a vibrational collective structure for these low-spin
excitations in $^{196,198}$Po.

To summarize, the energy level and branching ratio data support the
existence of vibrational collectivity in the Po isotopes which persists
to moderate angular momentum in $^{196,198}$Po.  To obtain a more microscopic
understanding of the onset of collective vibrational structure at
$^{198,196}$Po, we shall consider next the underlying single particle
structure.

\section{Single-particle interpretation}

Collectivity is a result of interactions between valence protons and
neutrons which occur when there is significant overlap between their
respective wavefunctions.  When this overlap is large, mixing occurs
which results in the lowering in energy of a coherent, collective state.
The yrast, low-lying levels in the semi-magic $^{210}$Po  nucleus are known to
consist predominantly of a simple ($\pi$h$_{9/2}$)$^2$ structure.
The neutrons are the hole orbitals given in Fig.~1 below the N=126
gap: p$_{1/2}$, p$_{3/2}$, f$_{5/2}$, and i$_{13/2}$. A necessary
condition for good overlap between the valence
protons and neutrons is that the angular behavior of the wavefunctions
is similar.

Schiffer~\cite{9,10}  developed a simple way to quantify semi-classically the
amount of angular overlap between single-particle wave functions.  When
two angular momentum vectors, j$_1$ and j$_2$, are added to form a resultant,
J, the three vectors obey the law of cosines, which can be written as

\begin{equation}
\theta_{j_{1},j_{2}, J} ~=~ \cos^{-1} \left\{  \frac{J^2 -
j_{1}^{2} - j_{2}^{2}}{2 |j_1 | ~ |j_2| } \right\}
\end{equation}
If the vectors are replaced with operators and the expectation value of
the operators are extracted then, for ${\rm j = j_1 = j_2}$

\begin{equation}
\theta_{j,J}  ~=~  \cos^{-1} \left\{ \frac{J (J+1) - 2 j
(j+1)}{2 j (j + 1) }  \right\}
\end{equation}

This ``semi-classical'' angle can be viewed as the angle two particles in
the same orbital, j, make with respect to one another in order to form
the total angular momentum, J, of a given state.  It gives a
quantitative measure of the distribution in angles of the nucleons in a
state with a particular J.

Table 5 presents the semi-classical angles of several $\mid$(j)$^2$,J$>$
configurations located near the Fermi surface in $^{196,198}$Po.  The
angles
for the neutron (i$_{13/2}$)$^{-2}$ and proton (h$_{9/2}$)$^2$
configurations up to
moderate angular momentum are similar.  This is consistent with our
assumption of collective motion that persists up to at least J$\approx$6
when the $\nu$i$_{13/2}$ orbital is near the Fermi surface.

The following then summarizes our understanding of the onset of
collectivity in Po nuclei as a function of neutron number and angular
momentum.  The $2_1^+$ level becomes collective as soon as there are valence
neutrons.  By N$\approx$122 with the opening of the 2f$_{5/2}$
orbital, the $4_1^+$ state
also has a predominantly 2-phonon collective character.  As N=112 is
approached, vacancies in the i$_{13/2}$ orbital occur, which allow
two-neutron configurations with J$>$4.  Given the sizable angular
overlap in
the ($\nu$i$_{13/2}$)$^{-2}$ and ($\pi$h$_{9/2}$)$^2$ wavefunctions,
collectivity persists beyond
J=4, with a predominantly collective $6_1^+$ state and, possibly less
collective, $8_1^+$ state.  Vibrational collectivity also characterizes the
low-spin non-yrast states in $^{196,198}$Po.

\section{Conclusion}

Evidence of collective vibrational motion that persists to moderate angular
momentum has been observed in the low-lying structure of $^{196}$Po.  The
energy spacings of the yrast 2$^+$, 4$^+$, 6$^+$ and, possibly, 8$^+$
states, as
well as of the $2_2^+$ and the $4_2^+$ levels,  are consistent with
the spacings of one, two,  three, and possibly four, phonon
multiplets.  The ratios
of the B(E2) values for decays from the $4_2^+$ and $2_2^+$ states are also
consistent with those of a vibrational structure.  The branching ratios,
together with the lack of a measureable intensity imbalance for the
$\Delta$J=0
transitions, indicate that it is unlikely that a 4p-2h proton
configuration plays a significant role in the non-yrast states.  The
transition at J $>$ 4 from single-particle behavior in $^{200}$Po to
collective
behavior in $^{196}$Po can be attributed to the opening of the
i$_{13/2}$ neutron
orbital, which allows neutron configurations of higher spin to interact
with the h$_{9/2}$ protons to form collective states.  A more quantitative
interpretation of these nuclei could come from a more detailed analysis
of the interaction between vibrational phonons and the two valence
protons, for example, within the particle-core coupling
model~\cite{11}.   Such
calculations are in progress~\cite{12}  and preliminary results confirm our
naive interpretation that $^{200}$Po has little collectivity for
J$>$4, $^{196}$Po
is an excellent multi-phonon vibrator, and that $^{198}$Po  is intermediate,
with wave functions which reflect a complicated interplay between
collective and two-proton configurations for J$>$4 excitations.

The present analysis has focused on relatively limited measures of
collectivity, i.e., energy levels and relative B(E2) values.  A
definitive measure of the collectivity as a function of neutron number
and angular momentum can come from absolute B(E2) values.  However,
these values are difficult to obtain from Doppler effects because of the
plethora of isomers in the level schemes of these polonium isotopes with
only a few valence particles.

\acknowledgments

We would like to thank Professor K. H. Maier for providing results on
$^{198}$Po  prior to publication.  This work has been funded in part by the
National Science Foundation (Rutgers) and the U.S. Department of Energy,
under contracts no. W-31-109-ENG-38 (ANL), W-7405-ENG-48 (LLNL), and
AC03-76SF00098 (LBL).

\begin{figure}
\caption{Energy levels for protons and neutrons in $^{208}$Pb, taken from
Ref.\ [3].}
\end{figure}

\begin{figure}
\caption{Coincidence spectrum gated on the 463-keV, $2 \rightarrow 0$
transition in
$^{196}$Po from the ANL measurements of the $^{172}$Yb($^{28}$Si,4n)
reaction at 141
and 145 MeV\@.  A multiplicity requirement K$\geq$5 was applied to the
coincidence matrix.}
\end{figure}

\begin{figure}
\caption{Level scheme of $^{196}$Po obtained from the present
measurements.  The widths of the arrows are proportional to the $\gamma$-ray
intensities.  The 11$^-$ isomeric state was taken from Ref.\ [1].  The
de-exciting transition from this isomer were not observed in the
present measurements, as indicated by the dotted line.  }
\end{figure}

\begin{figure}
\caption{Even mass polonium systematics from $^{210}$Po  (N=126) to $^{196}$Po
(N=112), taken from Refs.\ [1,6--8] and the present work.}
\end{figure}

\begin{table}
\caption{Energies, relative intensities and DCO ratios for the $\gamma$ rays
in $^{196}$Po from the $^{172}$Yb($^{28}$Si,4n) reaction measured at ANL.}
\begin{tabular}{dddc}
 Energy (keV)	&Intensity &	DCO ratio & Multipolarity\\
\tableline
237.28(9) & 29.0(4) & 1.26(35) &	E2\\
253.45(10)& 30.8(4) & 1.16(39) &	E2\\
387.64(12)& 46.8(6) & 0.75(25) & 	(E1)\\
395.82(12)& 13.1(4) & 1.44(31)\tablenotemark[1] &\\
414.24(8)& 44.8(8) & 0.88(33) &	E1\\
427.89(9)& 94.7(10)&  1.24(15) & 	E2\\
463.09(9)& $\equiv$100		& 1.41(14) & E2\\
485.79(10)& 18.9(4) &0.82(60)\tablenotemark[2] & (E1)\\
496.74(10)& 18.5(4)\tablenotemark[3] & & \\
499.11(10)& 69.7(7) &	1.48(26) &	E2\\
528.61(11)& 15.2(6)& 1.44(36) & 	E2\\
550.29(11)& 20.9(21) & 1.39(51) & 	E2\\
565.3(2)& 8.17(24) &	1.34(41)\tablenotemark[2] &	E2\\
583.9(2)& 51.7(7) & 1.96(54) & 	E2\\
616.9(2)& 43.7(8) &	1.68(24) &	E2\\
651.3(2)& 6.9(8)\tablenotemark[3] & \\
667.7(2)& 41.6(10) &	1.70(41) &	E2\\
859.2(2)& 20.5(7)  & 1.33(60)\tablenotemark[2] &	E2\\
911.5(3)& 20.1(6)\tablenotemark[3] & \\
\end{tabular}
\tablenotetext[1]{The LBL data indicates that the $\approx$396-keV
line is contamainated with another $\gamma$ ray that was not placed
in the level scheme.  Therefore, no multipolarity assignment was
made for the transition placed with the ANL data.}
\tablenotetext[2]{DCO ratio obtained from a sum gate on the 237 + 253
keV lines.}
\tablenotetext[3]{Insufficient intensity to extract the DCO ratio.}
\end{table}

\begin{table}
\caption{Intensities of transitions in the $4_2^+ \rightarrow 4_1^+
\rightarrow 2_1^+$ and $2_2^+ \rightarrow 2_1^+ \rightarrow 0$
cascades in $^{196,198}$Po.  In $^{196}$Po the intensities were
determined in
coincidence spectra gated on the 414 keV 5$^-$ $\rightarrow 4_2^+$ and
529 keV $4_2^+ \rightarrow 2_2^+$
transitions, respectively.  In $^{198}$Po the intensities were
determined in
similar spectra gated on the 391 keV $6_2^+ \rightarrow 4_2^+$ and 444
keV $4_2^+ \rightarrow 2_2^+$
transitions, respectively.}
\begin{tabular}{cccc}
   Nucleus &	Gate &	I$_{\gamma}$($2_2^+ \rightarrow 2_1^+$) &
I$_{\gamma}(2_1^+ \rightarrow 0_1^+$)\\
\tableline
$^{196}$Po &	529 keV	& 91(21) & 73(21)\\
$^{198}$Po  & 444 keV	& 89(31) & 100(31)\\
\tableline
Nucleus & Gate	& I$_{\gamma}(4_2^+ \rightarrow 4_1^+$) &
I$_{\gamma}(4_1^+ \rightarrow 2_1^+$)\\
\tableline
$^{196}$Po &	414 keV	& 59(22) & 44(24)\\
$^{198}$Po  & 391 keV	& 51(26) & 100(23)\\
\end{tabular}
\end{table}

\begin{table}
\caption{Energy ratios of yrast states in $^{196,198,200}$Po compared to
the theoretical predictions for (h$_{9/2}$)$^2$, vibrational, and rotational
models.  The (h$_{9/2}$)$^2$ predictions are for a surface delta
interaction.}
\begin{tabular}{cccc}
&	R(4/2) &	R(6/4) &	R(8/6)\\
\tableline
$^{200}$Po &	1.92	& 1.38 &	1.01\\
$^{198}$Po &	1.91	& 1.48 &	1.08\\
$^{196}$Po&  1.92	& 1.56 &	1.39\\
($\pi$h$_{9/2}$)$^2$ & 1.17 & 1.07 & 1.03\\
Vibrational & 2.00 &1.50 & 1.33\\
Rotational & 3.33 & 2.10 & 1.71\\
\end{tabular}
\end{table}

\begin{table}
\caption{Comparison between experimentally deduced E2 branching
ratios depopulating the non-yrast $2_2^+$ and $4_2^+$ states and theoretical
predictions for rotational, vibrational and 4p-2h models.  The
experimental transitions are assumed to be of pure E2 character; the
experimental errors given in parentheses are on the last digit(s).  The
rotational predictions assume K=0 for both quasibands.}
\begin{tabular}{cc|cccc}
&&\multicolumn{4}{c}{B(E2) Ratios}\\
Nucles & Transition & Experiment & Vibrator &	Rotor &	4p-2h\\
\tableline
$^{196}$Po & $4_2^+ \rightarrow 2_2^+$ / $4_2^+ \rightarrow 4_1^+$ &
1.0(5) & 1.10 &	$>$ 20 & $>$ 20\\
$^{196}$Po & $2_2^+ \rightarrow 2_1^+$ / $2_2^+ \rightarrow 0_1^+$ &
64(14) & $\infty$ & 2.04 (K=0)& Large E0\\
$^{198}$Po & $4_2^+ \rightarrow 2_2^+$ / $4_2^+ \rightarrow  4_1^+$ &
2.0(5) & 1.10 & $>$ 20 & $>$ 20 \\
$^{198}$Po & $2_2^+ \rightarrow 2_1^+$ / $2_2^+ \rightarrow 0_1^+$
&158(6)	& $\infty$  & 2.04 (K=0) & Large E0\\
\end{tabular}
\end{table}

\begin{table}
\caption{Values of semi-classical angles (in degree) of $\mid$(j)$^2$J$>$
configurations
for valence orbitals in polonium nuclei.  See text for details.}
\begin{tabular}{ccccc}
j (single particle) & J=2 & J=4 & J=6 & J=8\\
\tableline
i$_{13/2}$ & 160 & 144 &  127 & 108\\
h$_{9/2}$ & 152 & 127 & 99 & 63\\
f$_{5/2}$ & 131 &  82 & $-$ &$-$\\
p$_{3/2}$ &  90 & $-$ & $-$ &$-$ \\
\end{tabular}
\end{table}

\end{document}